\documentclass{emulateapj}

\usepackage{graphicx}

\def\ha{\relax \ifmmode {\rm H}\alpha\else H$\alpha$\fi}
\def\pa{\relax \ifmmode {\rm Pa}\alpha\else Pa$\alpha$\fi}
\def\arcsec{\hbox{$^{\prime\prime}$}}
\def\nii{\relax \ifmmode {\rm N\,{\sc ii}}\else N\,{\sc ii}\fi}
\def\hii{\relax \ifmmode {\rm H\,{\sc ii}}\else H\,{\sc ii}\fi}
\def\hi{\relax \ifmmode {\rm H\,{\sc i}}\else H\,{\sc i}\fi}
\def\deg{\hbox{$^{\circ}$}}

\begin{document}


\title{On the curvature of dust lanes in galactic bars}


\author{
S\'ebastien Comer\'on,\altaffilmark{1} Inma Mart\'inez-Valpuesta,\altaffilmark{1} Johan~H.~Knapen,\altaffilmark{1,2} and
John~E.~Beckman\altaffilmark{1,2,3}
}

\altaffiltext{1}{Instituto de Astrof\'isica de Canarias, E-38200 La
Laguna, Spain}

\altaffiltext{2}{Departamento de Astrof\'isica, Universidad de La Laguna, E-38205 La Laguna, Tenerife, Spain}

\altaffiltext{3}{Consejo Superior de Investigaciones Cient\'ificas, Spain}

\email{sebastien@iac.es,imv@iac.es,jhk@iac.es,jeb@iac.es}


\begin{abstract}

We test the theoretical prediction that the straightest dust lanes in bars are found in strongly barred galaxies, or more specifically, that the degree of curvature of the dust lanes is inversely proportional to the strength of the bar. The test used archival images of barred galaxies for which a reliable non-axisymmetric torque parameter ($Q_{\rm b}$) and the radius at which $Q_{\rm b}$ has been measured ($r(Q_{\rm b})$) have been published in the literature. Our results confirm the theoretical prediction but show a large spread that cannot be accounted for by measurement errors. We simulate 238 galaxies with different bar and bulge parameters in order to investigate the origin of the spread in the dust lane curvature versus $Q_{\rm b}$ relation. From these simulations, we conclude that the spread is greatly reduced when describing the bar strength as a linear combination of the bar parameters $Q_{\rm b}$ and the quotient of the major and minor axis of the bar, $a/b$. Thus we conclude that the dust lane curvature is predominantly determined by the parameters of the bar.

\end{abstract}




\section{Introduction}

Dust lanes in galactic bars have been observed in nearby galaxies for a long time. Photographic surveys from the beginning of 20$^{\rm th}$ century already allowed observation of these features. Pease (1917) describes the prominent and nearly straight dust lanes in NGC~5383 as a `dark streak'. The same paper presents images of NGC~1068 in which a dust lane is clearly visible. The dusty nature of these features was identified at a  much later stage. Sandage (1961) writes in his atlas that 'one of the major characteristics of the SBb(s) [galaxies] is the presence of two dust lanes leaving the nucleus, one on each side of the bar and extending into the spiral arms'.

Dust lanes have been recognised as related to shocks in the gas flow in barred galaxies since Prendergast (1962). Such shocks are usually found in the leading edge of the bar, roughly parallel to its major axis. The location of these shocks also corresponds to the location of areas of high shear, which prevents star formation (Athanassoula 1992; see Zurita et al.~2004 for a graphic illustration in the case of NGC~1530). Simulations show that the dust lanes are nearly straight and near the center of the bar when the galaxy has no inner Lindblad resonance (ILR), which is known to be quite a rare occurrence. ILRs cause the dust lanes to be offset from the bar major axis, and to be curved with the concavity pointing to the bar major axis (Athanassoula 1992). Simulations by Patsis \& Athanassoula (2000) show that the higher the gas sound speed, the smaller is the offset between the dust lane and the major axis of the bar. No dust lanes are expected in nuclear bars (Shlosman \& Heller 2002).

Athanassoula (1992) predicted from simulations that dust lanes would have a greater curvature in weaker bars and that dust lanes would be nearly straight in strong bars. This effect has been empirically verified using a set of images of nine barred spiral galaxies by Knapen et al.~(2002). The aim of the present Letter is to improve the statistics compared to the latter study, and to test definitively the prediction from Athanassoula (1992). Because dust lanes can be observed so easily they offer a fundamental handle on the underlying dynamics of the galaxy and the bar, yet they have hardly been studied observationally. A basic further aim of the present work is thus to explore in a statistical fashion the dependence of dust lane shape on the dynamical properties of the galaxy which hosts them.

\section{Sample selection and bar strength data}

We have selected all galaxies with a published non-axisymmetric maximum torque parameter for the bar, $Q_{\rm b}$, and a published radius at which $Q_{\rm b}$ has been measured, $r(Q_{\rm b})$. Its definition and the values for 266 different galaxies can be found in Laurikainen \& Salo (2002); Block et al.~(2004), Laurikainen el al.~(2004), Laurikainen et al.~(2006), and Comer\'on et al.~(2009). Some of these authors publish $Q_{\rm g}$, which includes the non-axisymmetries not related to the bars (but to spiral arms). However, $Q_{\rm g}$ and $Q_{\rm b}$ take very similar values in barred galaxies and can be interchanged for our purposes. In several cases $Q_{\rm b}$ has been reported in more than one paper, in which cases we have used the most recent measurement (except for those in Comer\'on et al.~2009 which are based on shallower images). For NGC~1068 we preferred the value from Comer\'on et al.~(2009) to that from Laurikainen et al.~(2004) because in the latter the authors measure $Q_{\rm b}$ of the nuclear bar, whereas the dust lanes are hosted by the large bar.

There are small variations in the techniques used to determine $Q_{\rm b}$ in the various articles quoted above but we have checked that the $Q_{\rm b}$ determinations are consistent within the error bars. These technical differences cause some scatter in the $Q_{\rm b}$ measurements. The most important source of scatter is the use of different vertical scale-heights when measuring $Q_{\rm b}$ and the large scatter in the observed vertical scale-height within one single Hubble type. For a detailed discussion see Laurikainen \& Salo (2002), where the authors consider that the uncertainties in both the orientation parameters and in the disc scale-height determination may cause an error of up to 10-15\% in $Q_{\rm b}$. As we are mixing data from several sources we take a slightly higher estimate for the overall uncertainty in our $Q_{\rm b}$ values of 20\%.

\begin{table*}

\caption{\label{sample} Properties of the host galaxies and dust lanes in our sample.}
\begin{center}
\begin{tabular}{l c c c c c c c| l c c c c c c c}

\hline
Galaxy &  Dist. & Disc PA & $\epsilon_{\rm disc}$ & $\Delta\alpha$  & $Q_{\rm b}$ & $r(Q_{\rm b})$ & Source & Galaxy &  Dist. & Disc PA & $\epsilon_{\rm disc}$ & $\Delta\alpha$  & $Q_{\rm b}$& $r(Q_{\rm b})$ & Source    \\
       &  (Mpc) & (\deg)&  &   (\deg)&            &    (\arcsec)         &             &        &   (Mpc)& (\deg) & & (\deg) &  & (\arcsec)&                                     \\
(1) & (2) & (3) & (4) & (5) & (6) & (7) & (8) & (1) & (2) & (3) & (4) & (5) & (6) & (7) & (8)\\
\hline

N150  & 20.2 & 107.6 & 0.502 &  10 & 0.459 &  26.7 & L04 &     N4314 & 16.4 & 61.8 & 0.041 & 74 & 0.442 &  52.5 & L04\\
N289  & 20.7 & 141.5 & 0.211 & 37 & 0.212 &  12.8   & L04 &     N4321 & 24.0 & 30.0 & 0.150 & 73 & 0.183 &  61.0 & L04\\
N613  & 18.7 & 121.6 & 0.228 &  0 & 0.401 &  68.4   & L04 &     N4457 & 13.4 &  80.8 & 0.117 & 37 & 0.089 & 31.5 & L04\\
N1068 & 15.3 & 70.0 & 0.150 & 111 & 0.094 &  55.0  & C09 &     N4548 & 16.4 & 153.2 & 0.256 & 59 & 0.344 &  55.5 & L04\\
N1084 & 18.5 & 52.7 & 0.247 & 35 & 0.212 &   35.5  & L04 &     N4579 & 23.0 & 94.8 & 0.217 & 50 & 0.197 &  34.5 & L04\\
N1097 & 15.2 & 130.0 & 0.320 & 60 & 0.279 &  75.0  & L04 &     N4593 & 35.3 &  98.1 & 0.258 &  28 & 0.309 &  45.5 & L04\\
N1300 & 20.2 & 150.0 & 0.240 & 63 & 0.537 &  68.4  & L04 &      N4651 & 13.0 &  73.1 & 0.388 & 29 & 0.120 &  16.5 & L04\\
N1365 & 20.1 & 32.0 & 0.450 &  64 & 0.490 & 164.0  & Bl04 &     N4691 & 16.3 & 41.2 & 0.158 & 11 & 0.504 &  13.5 & L04\\
N1433 & 11.6 & $-$  & 0.000 &  31 & 0.430 &  69.0  & Bl04 &    N5020 & 49.5 & 84.3 & 0.110 &  5 & 0.365 &  31.0 & C09\\
N1512 &  9.5 & 83.0 & 0.400 &  27 & 0.366 &  79.0  & C09 &    N5236 &   4.5 & $-$ & 0.000 &  41 & 0.294 &  81.0 & C09\\
N1530 & 38.9 & $-$  & 0.000 &  4 & 0.730 & 49.0   & Bl04 &       N5248 & 17.9 & 104.0 & 0.091 & 67 & 0.269 & 76.5 & L04\\
N1566 & 17.4 & 36.0 & 0.190 & 89 & 0.235 & 71.0   & C09 &        N5377 & 29.2 & 37.6 & 0.540 &  5 & 0.164 &  47.0 & C09\\
N1672 & 15.0 & $-$  & 0.000 & 60 & 0.349 & 59.0   & C09 &         N5457 &  6.9 & $-$  & 0.000 & 51 & 0.225 &  73.0 &L04\\
N2566 & 21.0 & 115.8 & 0.266 &  3 & 0.316 & 54.4  &  L04 &       N5643 & 14.8 & 131.2 & 0.102 & 12 & 0.415 &  33.6 &L04\\
N2964 & 20.6 & 96.8 & 0.434 &  20 & 0.310 & 22.5  & L04 &         N5728 & 39.7 & 14.5 & 0.410 &  27 & 0.350 &  57.0 &C09\\
N3166 & 18.8 &  82.5 & 0.414 &  25 & 0.107 & 31.5 & L04 &         N5806 & 20.5 & 174.3 & 0.480 & 55 & 0.140 &  41.0 &C09\\
N3184 & 14.5 & 135.0 & 0.050 & 95 & 0.160 &  45.0 & L02 &         N5905&  52.4 & 135.0 & 0.340 & 11 & 0.352 &  23.0 &C09\\
N3351 & 11.1 &   9.9 & 0.400 & 15 & 0.225 &  63.0 & C09 &         N5921 & 22.6 & 130.9 & 0.295 &  8 & 0.416 &  46.5 &L04\\
N3359 & 18.0 & 170.0 & 0.330 &  2 & 0.460 &  11.0 & L02 &         N5945 & 81.3 & 115.0 & 0.070 & 14 & 0.232 &  23.0 &C09\\
N3504 & 23.8 &  $-$  & 0.000 & 67 & 0.288 &  28.5 & L04 &         N6300 & 12.8 & 104.8 & 0.307 & 59 & 0.187 &  33.6 &L04\\
N3507 & 15.2 &  91.9 & 0.056 & 11 & 0.176 &  19.5 & L04 &         N6782 &  52.5 & 34.9 & 0.102 &  43 & 0.165 & 24.4 & L04\\
N3583 & 33.6 & 119.2 & 0.256 & 42 & 0.246 &  16.5 & L04 &         N6907 &  44.4 & 82.6 & 0.124 &  34 & 0.329 &  29.0 &L04\\
N3810 & 15.2 &  21.4 & 0.320 & 17 & 0.128 &  16.5 & L04 &         N6951 & 24.4 & 170.0 & 0.170 &  9 & 0.340 & 43.0 & Bl04\\
N3887 & 16.3 &  4.6 & 0.290 & 51  & 0.207 &  31.5 & L04 &         N7479 & 34.9 & 25.7 & 0.359 &  24 & 0.696 &  43.5 & L04\\
N4123 & 19.5 & 125.7 & 0.323 & 11 & 0.677 &  37.5 & L04 &         N7552 & 20.2 & 169.9 & 0.127 &  14 & 0.395 &  45.2 &L04\\
N4212 & 16.8 &  75.7 & 0.337 & 70 & 0.234 &  28.5 & L04 &         N7582 & 20.2 & 152.9 & 0.529 &  1 & 0.436 &  56.8 & L04\\
N4274 & 15.6 & 102.0 & 0.500 & 36 & 0.310 &  45.0 & L02 &         N7723 & 25.2 & 38.5 & 0.307 &  1 & 0.349 &  16.5 &L04\\
N4303 & 23.1 & 146.9 & 0.139 & 65 & 0.259 &  40.5 & L04 \\

\hline
\end{tabular}
\end{center}
Notes: galaxy ID (col.~1), distance to the galaxy in Mpc from Comer\'on et al.~2009 and LEDA (col.~2), galaxy disc PA (col.~3) and disc ellipticity from the literature (col.~4), dust lane curvature as defined in Sect.~3 (col.~5), $Q_{\rm b}$ (col.~6), $r(Q_{\rm b})$ (col.~7), and literature source for the galaxy disc PA, ellipticity, $Q_{\rm b}$, and $r(Q_{\rm b})$ (col.~8). L02: Laurikainen \& Salo (2002), Bl04: Block et al.~(2004), L04: Laurikainen et al.~(2004), and C09: Comer\'on et al.~(2009).

\end{table*}

The 266 galaxies with a $Q_{b}$ determination in the literature have morphological types spanning from S0/a to Sd. For those galaxies in the Sloan Digital Sky Survey (SDSS) DR7 (Abazajian et al.~2009) we examined single-band ($g$ and $i$) and color-index ($g-i$) images to select galaxies presenting clear dust lanes. For all other galaxies we examined all optical images available in the NASA-IPAC Extragalactic Database (NED). Out of the original 266 galaxies 55 have been found to host dust lanes. This is a lower limit to the true proportion of recognizable bar dust lanes in disc galaxies because for several galaxies only Digital Sky Survey (DSS) images are available, on which bars are often saturated. The final sample for our dust lane study is thus of 55 galaxies. Table~\ref{sample} lists these galaxies, their $Q_{\rm b}$, and its source.

\section{Measurement of the dust lane curvature}

The curvature of the dust lanes was determined directly from the SDSS and NED images using the procedure described by Knapen et al.~(2002). We first deprojected the galaxy images using the deprojection parameters listed in Table~\ref{sample}. We then measured the change in the angle of the tangent to the dust lane in the range in which its curvature is constant, thus often ignoring the inner parts of the dust lanes where $x_{2}$ orbits in the circumnuclear region influence their shape and enhance the curvature (Athanassoula 1992). We also often ignored the outermost parts of the dust lanes, where they are closer to the corotation radius. The measure of dust lane curvature, $\Delta\alpha$, as used here is arrived at by multiplying the change in the tangent angle defined above by the only measure of the bar length that can be reliably derived in both observed and modelled bars, namely the radius at which the torque is maximal ($r(Q_{\rm b})$). $\Delta\alpha$ is thus a dimensionless quantity which takes into account the size of the host bar. The results are listed in Table~\ref{sample}.

The uncertainty in the dust lane curvature is hard to quantify. The error due to galaxy deprojection is most probably not very large ($<10\deg$) because in most cases dust lanes are intrinsically rather straight, and thus less sensitive to deprojection than a more strongly curved feature might be. The uncertainty in determining the tangent to the dust lane is larger because the dust lanes are often not too well defined, and sometimes rather fluffy. By repeating our measurements for the same dust lanes, but using slightly different tangent points, we found that the results agree to within a few degrees per kpc. We thus adopt a conservative estimate of uncertainty in the dust lane curvature of  $\delta\Delta\alpha\sim15\deg$.

\section{Results}

\begin{figure}
\begin{center}
\begin{tabular}{c}
\includegraphics[width=0.45\textwidth]{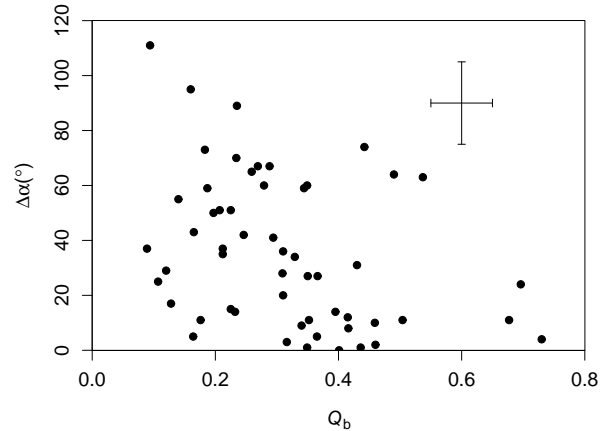}\\
\end{tabular}
\caption{\label{qbs} Dust lane curvature related to $Q_{\rm b}$.}
\end{center}
\end{figure}

The relation between the bar strength as described by $Q_{\rm b}$, and the dust lane curvature, $\Delta\alpha$, is shown in Fig.~\ref{qbs}. The plot has the axes flipped with respect to those used in Athanassoula (2002) and Knapen et al.~(2002). The most striking feature in the plot is the clear division between a populated lower left and an unpopulated upper right part, separated by  a straight and well-defined envelope. This implies that strong bars cannot have curved dust lanes, and that galaxies with smaller bar strengths can have progressively more curved dust lanes. A small unpopulated area appears in the extreme lower left, at the location of galaxies with straight dust lanes and small $Q_{\rm b}$. If on the vertical axis of the plot we use the change in the angle of the tangent to the dust lane (in \deg/kpc) instead of the dust lane curvature the results remain essentially the same.

A similar diagram was presented by Knapen et al.~(2002), but they only plotted nine data points. In their figure, the points appear roughly aligned, but in our new version, and thanks to the much larger number of data points, we can see how the relationship has a spread much larger than can be expected from the error bars. It is thus reasonable to suppose that although $Q_{\rm b}$ limits the allowed dust lane curvature, it does not prescribe it. Other factors must determine the degree of curvature, especially for the weaker bars, and we used numerical simulations to explore which factors might contribute.

\section{Simulating dust lanes in barred galaxies}

We ran a set of 238 3D numerical simulations and studied the dynamical response of the gas within a barred potential using the code FTM~4.4 (updated version) from Heller \& Shlosman (1994). In each individual simulation, we used $10^5$ isothermal, non-self-gravitating, collisional gas particles. We reproduced the potential of barred galaxies by means of a Miyamoto-Nagai disk, bulge and halo. The bar is reproduced by a Ferrers potential (Ferrers 1877), which is introduced gradually, in one rotation, in a way that conserves the mass.

We use different parameters for the bar ellipticity, mass, and pattern speed, and for the bulge mass ratio (B/B+D) in order to cover the complete range of galaxies in our set of observations presenting dust lanes.

The Miyamoto-Nagai potential is described by the equation:

\begin{equation}
\begin{centering}
\Phi_{\rm M}(R,z)=-\frac{GM}{\sqrt{R^2+(a+\sqrt{z^2+b^2})^2}}
\end{centering}
\label{eqn:MiyNag}
\end{equation}

For the disk we use $M_{\rm d}=5\times10^{10}\,M_{\sun}$, $a=5.0$~kpc and $b=0.5$~kpc. For the bulge we use $b=0.05$ varying B/B+D from $0.1$ to $0.5$ in steps of $0.1$ by increasing the bulge mass. For the halo we use $M_{\rm h}=10^{11}\,M_{\sun}$, $a=0.0$~kpc, and $b=10$~kpc.

The Ferrers potential is described by

\begin{equation}
\begin{centering}
\rho=\frac{15}{8}\frac{M_{\rm b}}{\pi abc}(1-m^2)\,\,\,{\rm for}\,\,\,m\le1,\rho=0\,\,\,{\rm for}\,\,\,m>1
\end{centering}
\end{equation}

where $m^2=\frac{x^2}{a^2}+\frac{y^2}{b^2}+\frac{z^2}{c^2}$, and $a$ is the major, $b$ the minor, and $c$ the vertical axis, taken here in such a way that $a/b$ takes the values $1.5,2.0,2.5,3.0,3.5,4.0,4.5$ accounting for the ellipticity of the bar. We use $a=6$\,kpc, $c=b/4$ and two different values for the bar mass: $M_{\rm b}=5.3\times10^{10}\,M_{\sun}$ and $M_{\rm b}=8\times10^{10}\,M_{\sun}$.

The pattern speed is chosen so that in a linear approximation the corotation radius is always greater than the length of the bar. In our set of simulations this means $10$, $20$, $30$, and $40$~km\,sec$^{-1}$\,kpc$^{-1}$.

\begin{figure*}
\begin{center}
\begin{tabular}{c}
\includegraphics[width=0.45\textwidth]{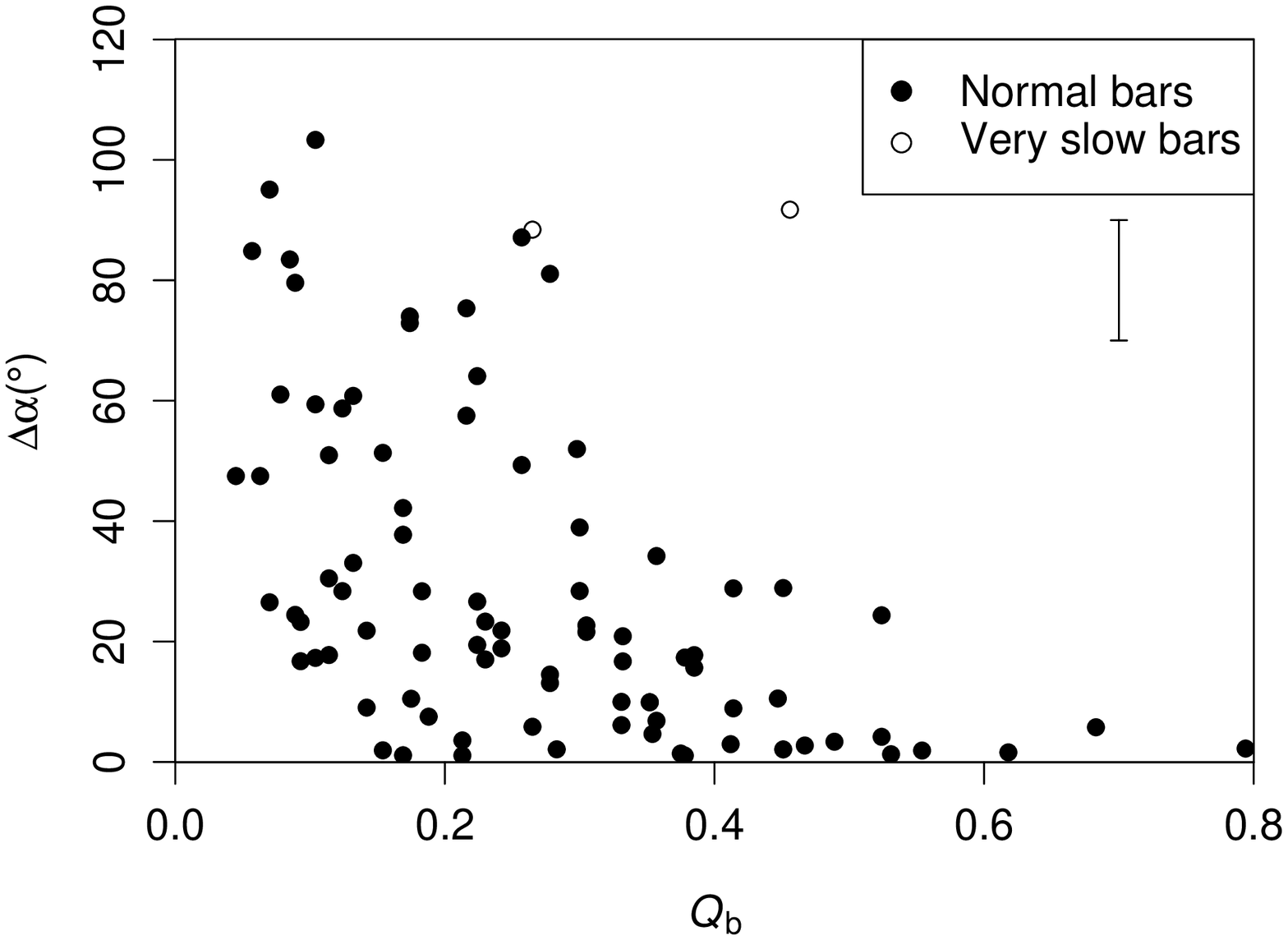}
\includegraphics[width=0.45\textwidth]{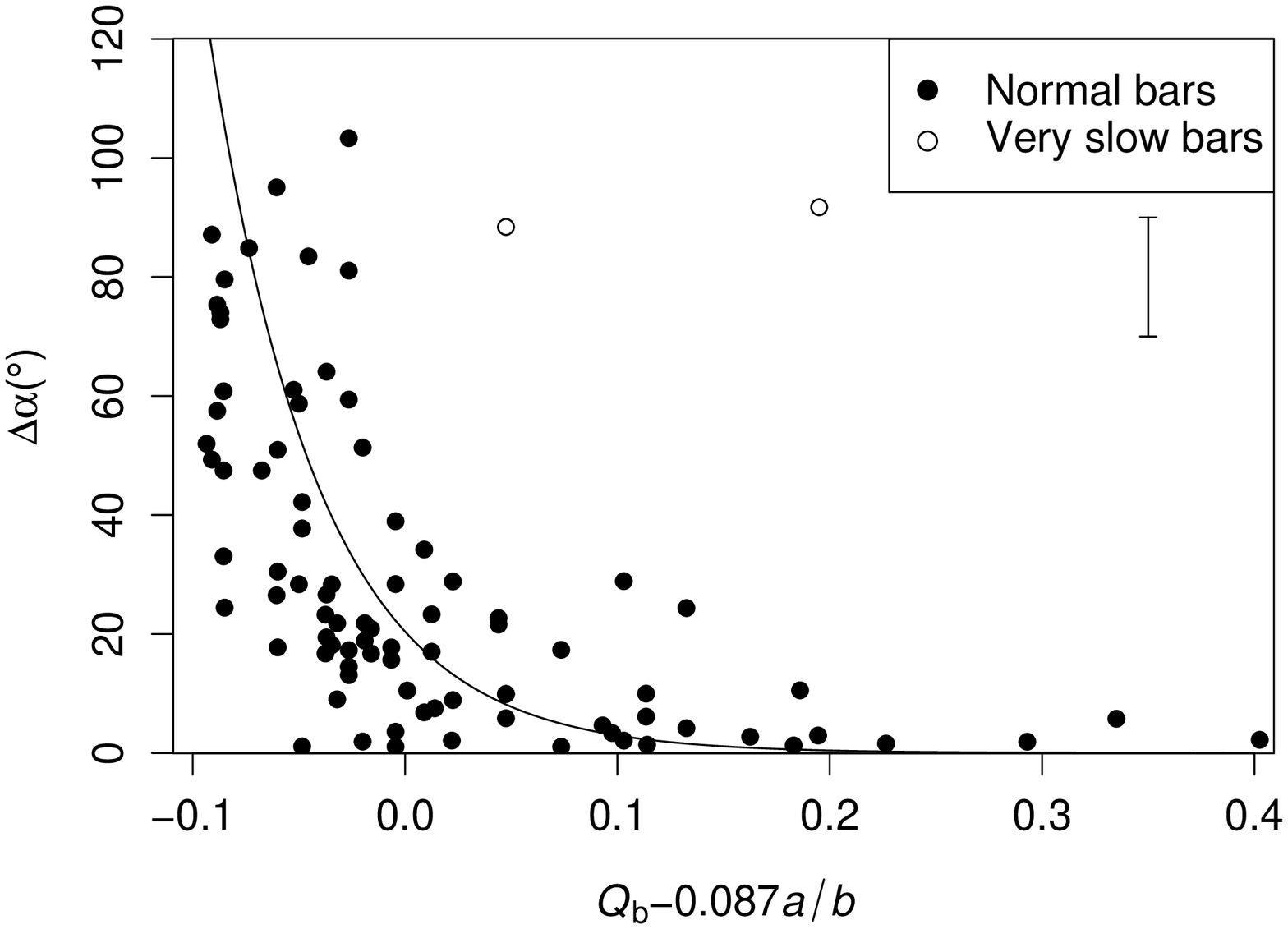}\\
\end{tabular}
\caption{\label{simul} Left panel: dust lane curvature related to $Q_{\rm b}$ in the simulated galaxies. Filled dots stand for `normal' bars ($r_{\rm CR} < 3.0\,r_{\rm b}$) and empty dots stand for `very slow' bars ($r_{\rm CR} > 3.0\,r_{\rm b}$). Right panel: dust lane curvature related to a combination of $Q_{\rm b}$ and $a/b$ which minimizes the spread. The line is the best fit to the `normal' bar data. There is no error bar on $Q_{\rm}$ because it has been analiticaly calculated from the potentials used in simulations.}
\end{center}
\end{figure*}

As dust lanes are found in areas with gas shocks (Prendergast 1962), we inferred dust lane curvature in the simulated galaxies by measuring the curvature of gas density enhancements using the same methodology as used for the dust lanes in real galaxies. This measurement was applied after two bar rotations. Well defined measureable dust lanes appeared in 88 out of 238 simulations. The results can be seen in the left panel of Fig.~\ref{simul}, where the symbol coding indicates the ratio between the corotation radius, $r_{\rm CR}$, and the bar radius, $r_{\rm b}$. For convenience, we define in the context of this Letter that for $r_{\rm CR} < 3.0\,r_{\rm b}$ we have a `normal' bar and for $3.0\,r_{\rm b} < r_{\rm CR}$ we have a `very slow' bar (more details on this bar classification can be found in Debattista \& Sellwood 2000). $r_{\rm CR}$ was always measured using a linear approximation and $r_{\rm b}=6$\,kpc was fixed in the simulations. `Very slow' bars (two out of 88 in the simulations which present dust lanes) are rare in nature (just one is found in the compilation by Rautiainen et al.~2008) and thus are not included in the discussion in the next section because they have a non-standard behaviour.

\section{Discussion}

The observational part of our study confirms that the degree of curvature of dust lanes decreases with increasing bar strength, i.e.~stronger bars have straighter dust lanes (Fig.~\ref{qbs}). This was predicted by Athanassoula (1992) and confirmed in a preliminary way from observations of nine galaxies by Knapen et al.~(2002). Here, we find that the relation, in fact, outlines an upper limit to the dust lane curvature allowed for a particular value of $Q_{\rm b}$, in the sense that a bar with low $Q_{\rm b}$ can have either straight or curved dust lanes, but bars with high $Q_{\rm b}$ can only have straight dust lanes.

We performed a series of numerical simulations to explore which parameters may be at the origin of the spread of the data points in the left panel of Fig.~\ref{simul}. We tested whether this spread is caused by different bulge masses, different galaxy masses inside the corotation, and different bar pattern speeds. We find no effect of these parameters on the spread except for the bar pattern speed: the spread for slow bars is higher but this effect is quite small. We found, however, that the data can be organized in a plot with much smaller spread by fitting the logarithm of the dust lane curvature to a linear combination of $Q_{\rm b}$ and $a/b$ which are parameters that depend only on bar properties. The best fit is
\begin{equation}
Q_{\rm b}-0.087\,a/b=(0.156\pm0.020)-(0.119\pm0.015)\,{\rm log}_{10}\Delta\alpha
\end{equation}
and it is represented in the right panel of Fig.~\ref{simul}. The correlation coefficient of the fit is $\rho=0.66$, better than when just fitting $Q_{\rm b}$ and ${\rm log}_{10}\Delta\alpha$ ($\rho=0.58$). This fit shows that $Q_{\rm b}$ overestimates the effect of the bar ellipticity on the dust lane curvature.

The correlation between the linear combination of $Q_{\rm b}$ and $a/b$ and the logarithm of $\Delta\alpha$ cannot yet be easily tested with observational data. Measuring the ellipticity of the bar requires an accurate decomposition of high quality images of the galaxy into its different components. Further work is needed in order to obtain these decompositions and refine the comparison with the simulations.

\section{Conclusions}

Using 55 galaxies with published measurements of the bar strength indicator $Q_{\rm b}$, and for which we could obtain accurate measurements of the dust lane curvature, $\Delta\alpha$, we confirm the theoretical prediction made by Athanassoula (1992) that the dust lane curvature anticorrelates with the bar strength. Strong bars thus host straight dust lanes, while weaker bars can
host more curved dust lanes. We do find that the anticorrelation has a
large spread, but from a set of 238 numerical simulations of barred
galaxies we show that this spread can be greatly reduced by using an
appropriate linear combination of bar parameters ($Q_{\rm b}$ and the axis ratio
$a/b$).

%===============================================================================

\section*{Acknowledgments}

We acknowledge Herv\'e Bouy for his help with statistics software. Support by the Ministerio de Educaci\'on y Ciencia (AYA 2004-08251-CO2-01 and AYA 2007-CO2-01), the Instituto de Astrof\'\i sica de Canarias (P3/86 and 3I2407) is gratefully acknowledged. This research has made use of the NASA/IPAC Extragalactic Database (NED) which is operated by the Jet Propulsion Laboratory, California Institute of Technology, under contract with the National Aeronautics and Space Administration. We acknowledge the usage of the HyperLeda database (http://leda.univ-lyon1.fr). Funding for the SDSS and SDSS-II has been provided by the Alfred P. Sloan Foundation, the Participating Institutions, the National Science Foundation, the U.S. Department of Energy, the National Aeronautics and Space Administration, the Japanese Monbukagakusho, the Max Planck Society, and the Higher Education Funding Council for England. The SDSS Web Site is http://www.sdss.org/.

%===============================================================================

\end{document}